\documentclass{article}

\usepackage{arxiv}

\usepackage[utf8]{inputenc} 
\usepackage[T1]{fontenc}    
\usepackage{hyperref}       
\usepackage{url}            
\usepackage{booktabs}       
\usepackage{amsfonts}       
\usepackage{nicefrac}       
\usepackage{microtype}      
\usepackage{lipsum}
\usepackage{amsmath}
\usepackage{graphicx}
\graphicspath{ {./} }
\title{Discovering novel cancer bio-markers in acquired lapatinib resistance using Bayesian methods}

\author{
  AKM Azad \\
  iThree Institute \\ Faculty of Science \\ University of Technology Sydney, Australia \\
  \texttt{akm.azad@uts.edu.au} \\
   \And
 Salem Alyami\thanks{corresponding author} \\
  Department of Mathematics $\&$ Statistics \\ Imam Mohammad Ibn Saud Islamic University\\ Riyadh, Saudi Arabia \\
  \texttt{saalyami@imamu.edu.sa} \\
}

\begin{document}
\maketitle

\begin{abstract}
Genes/Proteins doesn’t work alone within our body, rather as a group they perform certain activities indicated as pathways. Signalling transduction pathways (STPs) are some of the important pathways that transmits biological signals from protein-to-protein controlling several cellular activities. However, many diseases such as cancer target some of these signalling pathways for their growth and malignance, but demystifying their underlying mechanisms are very complicated tasks. In this study, we use a fully Bayesian approaches to develop methodologies in discovering novel driver bio-markers in aberrant STPs given two-conditional high-throughput gene expression data. This project, namely PathTurbEr (Pathway Perturbation Driver), is applied on a global gene expression dataset derived from the lapatinib (an EGFR/HER dual inhibitor) sensitive and resistant samples from breast cancer cell-lines (SKBR3). Differential expression analysis revealed 512 differentially expressed genes (DEGs)  and their signalling pathway enrichment analysis revealed 22 singalling pathways as aberrated including PI3K-AKT, Hippo, Chemokine, and TGF-$\beta$ singalling pathway as highly dys-regulated in lapatinib resistance. Next, we model the aberrant activities in TGF-$\beta$ STP as a causal Bayesian network (BN) from given observational datasets using three Markov Chain Monte Carlo (MCMC) sampling methods, i.e. Neighbourhood sampler (NS) and Hit-and-Run (HAR) sampler, which has already proven to have more robust inference with lower chances of getting stuck at \textit{local optima} and faster convergence compared to other state-of-arts methods. Next, we examined the structural features of the optimal BN as a statistical process that generate the global structure using, $p_1$-model, a special class of Exponential Random Graph Models (ERGMs) and MCMC methods for their hyper-parameter sampling. This step will enable us to detect key players that would supposedly drive the aberration within the chosen BN structure of STP, which yielded 31, 33, and 22 perturbation driver genes out of 79 constituent genes of three STP models of TGF-$\beta$ signalling. Functional enrichment with GO terms of these driver genes suggested their significant associations with breast cancer progression/resistance. The full code base for \textbf{PathTurbEr} is available in here: \url{https://github.com/Akmazad/PathTurbEr/}.
\end{abstract}


\section{Introduction}
Signal transduction pathways (STPs) involve collections of signalling proteins, transducing biological signal from one to another that controls cell growth and division, cell death, cell fate, and cell motility \cite{Sever2015}. Hence, altered activities in one or more of these STPs, e.g. PI3K/AKT signalling, Ras/MAP signalling, Notch signalling and Wnt signalling mediates uncontrolled cell growth/death resulting tumour initiation, progression, metastasis \cite{Grumolato2017}, or even resistance phenotype to targeted therapies \cite{Azad2017}. Induction of these perturbed signalling activities often refers to oncogenic mutations of key signalling proteins that are structurally central (hub) in corresponding STP structures, resulting in their or others’ hyper-activation or inactivation \cite{Sever2015}. Therefore, elucidating mechanisms underpinning aberrant signalling activities are crucial for successful cancer therapeutics.

Hight-throughput datasets e.g. Gene/Protein expressions, copy number variation, methylation, and microRNA expression from cohorts of cancer-related sample patients (i.e. treated, untreated, and resistant) enable us genome-wide analyses of disease phenotypes (e.g. cancer malignance or cancer drug-resistance) and identification of relevant biomarkers in silico. Utilizing those high-throughput datasets for deriving data-driven models facilitates critical assessments of the systems behaviour in response to any perturbation from wild-types, i.e. due to the presence of disease markers, treatment with drugs or resistance phenotypes. Therefore, these data-driven model inference offers unique scope to deduce novel findings that may better reflect the systems dynamics under study. However, it’s a very complex computational task and still an open problem in the field of systems biology since inferring optimal model from the data require smart searching algorithms within the space of all possible models. The simplest of all approaches could probably be deriving co-expression network from available datasets among all signalling proteins but those models will lack causalities that are often crucial for better understanding interactions among biological entities.

A BN is defined as a directed acyclic graph (DAG), `G' = (V,E), where `V' are a set of random variables (here genes/Proteins) and ‘E’ represents relationships among those random variables. Each random variable is associated with conditional probability distributions given it’s parents except the root variable, which are only associated with corresponding prior probability distributions. Causal BNs possess causal edges, for example X$\longrightarrow$Y indicates causality from random variable `X' to random variable `Y'. An important property of BNs is called ‘Markov condition’, which must be satisfied for the probability distributions of the constituent random variables. ‘Markov condition’ states that each variable in the BN should be conditionally independent of other random variables that are its non-descendants conditional on its parents.

To model STP as a BN, we can model the phosphorylation activities of signalling proteins as random variables and any phosphorylation activities (i.e. signalling activities) among signalling proteins as arcs among those variables. Nonetheless, BN models like such must meet the above-mentioned `Markov condition' to represent the joint probability distribution of the random variables. Woolf \textit{et al.} \cite{Woolf2005} suggested that the steady-state concentration of signalling proteins and their corresponding signalling activities supports this condition, and Sachs \textit{et al.} \cite{sachs2006} has conducted a successful proof-of-principle analysis on that. 

Recently, Bayesian network (BN) has been applied to model aberrant STPs by studying case/control gene expression data \cite{Neapolitan2014} using MCMC methods such as Metropolis-Hastings or Gibbs sampling algorithm. Like other non-MCMC approaches (i.e. score-based methods such as Greedy, Hill-Climbing, Tabu search algorithm), these MCMC approaches also suffers local optimal problem, which means models claimed as optimal from those methods may not be globally optimal. Recently, we have shown that compared to Metropolis-Hastings algorithm, newly developed Neighbourhood sampler (NS) and Hit-and-Run sampler finds BN structures with better accuracy and faster convergence towards true distribution. For example, the main rationale behind NS sampler’s efficiency lies in its reduction step, where the rejected BNs are excluded from being chosen second time, offering greater chances to others to be selected, and therefore have less chance to get stuck at local optima \cite{NSpaper2016}. We hope using these two methods, we will be able to infer robust BN as a perturbed STP in given high-throughput datasets.

Biological networks reveal scale-free topology where majority of nodes possess fewer connections whereas small of amount nodes have very high connectivity, typically known as hub nodes \cite{barabasi204}. It is shown that data-driven network models also feature the similar behaviour \cite{Ernst2017}, i.e. the existence of hub nodes within the network. This behaviour is expected in biological systems as for disease perturbations to take place and/or spread within the system, it must target central players that has maximum reachability to other nodes via excessive connectedness. Hence hub nodes serve as key bio-markers which must be analysed, found and targeted in order to counteract the perturbations. 
It’s important to acknowledge that biological processes (e.g. signalling activities) underlying data and the experimental measurements are stochastic. For example, derived interactions in the data-driven BN model may not be truly reliable whereas some critical interactions may remain undetected. Therefore, after modelling data-driven STP (optimal BN), statistical approaches relying on probabilistic models should be adopted to assess the probability of forming each interactions (i.e. aberrant activities among signalling proteins) within that BN structure. Being originally proposed in social network analyses, Exponential Random Graph Models (ERMGs), or p*-models \cite{wasserman_robins_2005} are central of statistical modelling of networks, where each possible edge in a network is considered as a random variable and modelled as combinations structural properties of the constituent nodes, e.g. global density of nodes, attractiveness/expansiveness of nodes (commonly known as sociality parameters). Note, non-statistical methods like descriptive approaches may also model hub nodes based on their degree distributions which may not address stochastic nature of the underlying data and experimental measurements \cite{Bulashevska2010}. In the context of this study, hub nodes should be highly social, which would allow us to analyse the possibilities of detecting those hub nodes as a key bio-marker underlying the aberrant STP structure formation. 

The primary research questions we would like address in this study are as two-folds. First, \textit{is there exists a causal structure that optimally model the aberrant signalling activities within a signal transduction pathways (STP) given two-conditional datasets e.g. cancer-vs-normal or resistant-vs-treatment?} Second, \textit{how the data-driven STP structure has emerged – is there any local biological process or local structure (e.g. star-like shapes) that contributed in generating the global structure?} We hypothesize that the capabilities of new MCMC sampling methods in accurate BN inference can be leveraged for finding optimally aberrant STP structures by studying two conditional studies – structures that would model perturbed signalling activities in cancer-vs-normal or resistant-vs-treatment conditions. In statistical modelling, it can be hypothesized that global structure of a network emerges from the agglomeration of local structures that relies on the local inter-dependencies of edges. We also hypothesize that statistical analyses of the generic and structural properties of the inferred aberrant STP using ERGM models along with hierarchical Bayesian modelling of their hyper-parameter inter-dependencies would unravel key genes/proteins (e.g. network hubs) that drives those perturbations within that particular STP.

\section{Methods}
\begin{figure}[hbt!] 
\centering
\includegraphics[scale=0.35]{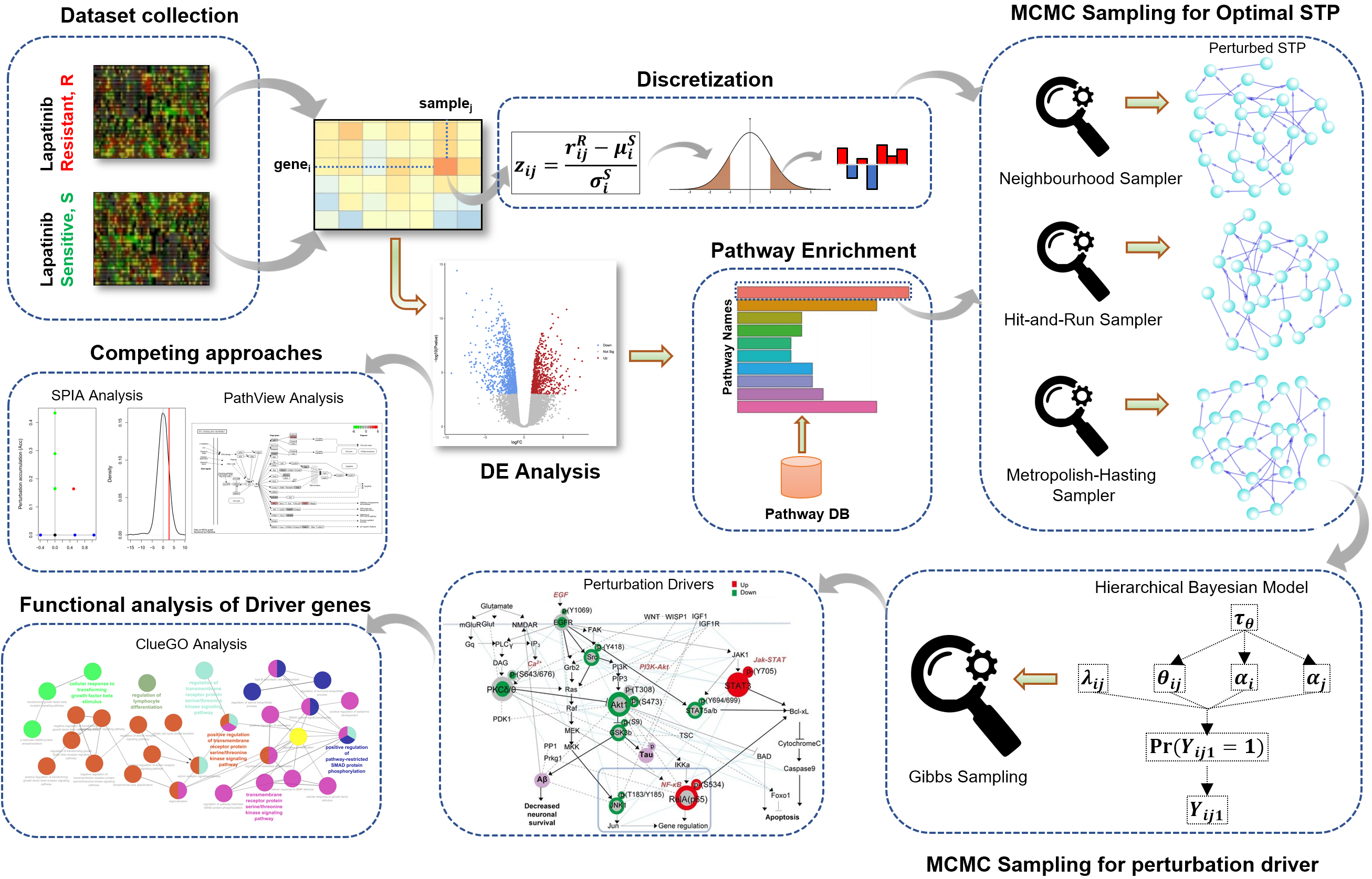}
\caption{\textbf{PathTurbEr}:
      a schematic diagram of our proposed approach.}
\label{fig_schematic} 
\end{figure}

\textit{PathTerbEr} is a data-driven approach in detecting perturbation driver bio-markers in signaling pathways. The schematic view of the proposed research plan is depicted in Figure ~\ref{fig_schematic}.

\subsection*{\textbf{Acquiring datasets and pre-processing}}
A global gene expression dataset for lapatinib sensitive-vs-resistant samples, for breast cancer cell-line, namely SKBR3, from Gene expression omnibus (GEO) with accession number GSE38376, published by Komurov \textit{et al.} \cite{Komurov2012}. This dataset includes samples with basal conditions (i.e. lapatinib-sensitive) dosed with 0$\mu$M, 0.1$\mu$M, and 1.0$\mu$M of lapatinib drug, and samples with resistant conditions with the same doses, respectively. Thus, two data matrices, each for lapatinib-sensitive and lapatinib-resistant were obtained, which was further considered as control-vs-case data for the downstream analyses. 45 signalling pathways were collected from KEGG pathway database \cite{KEGG2002}. 

\subsection*{\textbf{Differential expression and functional enrichment test}}
For the differential expression analysis, the raw gene expression were fitted with linear models using limma R package \cite{limma2015} with eBayes function to calculate the Bayes statistics. Furthermore, the significance score (i.e. p-values) were further adjusted with FDR correction techniques. Next, a set of differentially expressed genes (DEGs) based on the adjusted p-values, were used for over-representation test with 45 KEGG signalling pathways using the hyper-geometric test by \textit{phyper} function available in stats R package. The enrichment significance score (p-values) were further adjusted using FDR-correction.

\subsection*{\textbf{Inferring optimally perturbed STP structure using MCMC sampling algorithms}}
In this study, we have modelled the signal transduction pathways (STPs) as Bayesian Networks (BNs), which is a graphical model for inferring causal dependencies among statistical variables, which is the genes/proteins in this case. Let, $G:= (V,E)$ be a Bayesian network as graph with V is the set of genes, and E is the set of causal edges (i.e. directed) connecting the nodes from V, which are also acyclic and connected graph. 

\subsubsection*{\textbf{Parameter inference of BNs using Dirichlet-Multinomial distribution}}
Prior to BN structure learning, we modeled the perturbation of the lapatinib-resistant samples using the following equation:

\begin{equation}
    z_{ij}^{R} =  \frac{r_{ij}^{R} - \mu_i^{S}}{\sigma_i^{S}} 
\end{equation}
where, $r_{ij}^{R}$, $\mu_i^{S}$, $\sigma_i^{S}$, and $z_{ij}^{R}$ are raw expression value of a $gene_i$ in the $sample_j$, the average and standard deviation of all the samples in the lapatinib-sensitive condition of $gene_i$, and the transformed value of $R_{ij}$, respectively. Next, the transformed lapatinib-resistant data matrix were further discretized using the following formula:

\begin{equation}
  z_{ij}^{R\prime} = 
 \begin{cases}1 & z_{ij}^{R} > 1.5 \\-1 & z_{ij}^{R} < -1.5\\0 & otherwise\end{cases}  
\end{equation}

In BNs, the causality is modelled as the conditional probabilities between parent and the child nodes, where the children nodes probabilities are conditioned on the probabilities of parents. With a given data $D$ and a candidate BN structure, $G$, a conditional probability is defined as $P(X_i|Pa(X_i))$, where $X_i$ is a particular gene, and $Pa(X_i)$ is the set of parents of $X_i$. When the data, $D$ is given, this conditional probability can be defined as $P(X_i=k|Pa(X_i)=j) = {\theta}_{ijk}$, where ${\theta}_ijk$ is is the probability of each state value $k$ within each node $X_i$, given that its parents are in configuration $j$. Finally, to calculate the likelihood score (i.e. probability) of a BN structure, $G$, a multinomial distribution (MD) was used to relate the data with model, with dirichlet distribution for its prior. Hence, the formula for the Dirichlet-Multinomial model is as follows:

\begin{equation}
P(D|G) = \prod_{i=1}^n\prod_{j=1}^{q_i} \frac{\Gamma(\alpha_{ij})}{\Gamma(\alpha_{ij}+N_{ij})} \prod_{k=1}^{r_i} \frac{\Gamma(\alpha_{ijk}+N_{ijk})}{\Gamma(\alpha_{ijk})}
\label{mul}
\end{equation}
where $N_{ijk}$ is the number of observations in bin $k$ of node $i$ corresponding to a parent configuration $j$, $r_i$ is the number of possible state values (bins) for a particular variable $X_i$, $q_i$ is the total number of configurations of parent state values of $X_i$, $N_{ij}=\sum_{k=1}^{r_i} N_{ijk}$, and $\alpha_{ijk}$ are the hyper-parameters (hyper-conditional probabilities), $\alpha_{ij}=\sum_{k=1}^{r_i} \alpha_{ijk}$. Here, we have considered $\alpha_{ijk}=\frac{\alpha}{q_i r_i}$, where $\alpha$ is the total imaginary counts for the Dirichlet prior. The posterior probability distribution of the graph $G$ given data $D$ can now be constructed as:
\begin{equation}
P(G|D) = \frac{P(D|G) P(G)}{P(D)}   
\label{likelihood}
\end{equation}
where we only consider the numerator as the $P(D)$ does not depends of $G$. Moreover, we've used a uniform prior for $P(G)$. 

\subsubsection*{\textbf{Structure inference using Neighbourhood sampler, Hit-and-Run and Metropolis-Hastings sampler}}

Here, we have three MCMC (Markov Chain Monte Carlo) sampling algorithms, namely Neighbourhood sampler, Hit-and-Run sampler and Metropolis-Hastings algorithms for inferring optimal STP from the whole BN search space. At each sampling iteration, each of these algorithms considers a candidate BN and looks around its neighbourhood space and evaluates the likelihood of it using equation \ref{likelihood}, and selects the one that maximizes the likelihoods as optimal BN models for a signalling pathway of interest. The details of each of these algorithms are explained in the original articles \cite{NSpaper2016,Salemthesis2017}, which is beyond the scope of this article. Finally, the inferred BN STPs (from each of these MCMC sampling algorithms) was considered as the optimally perturbed STP structure model.

\subsection*{\textbf{Bayesian statistical modelling of the optimal STP to infer perturbation driver genes}}

\subsubsection*{\textbf{Network model}}
In this study, we have used a fully Bayesian statistical modelling approach for analysing the statistical aspect of the perturbed STP structure yielded from the MCMC sampling algorithms of BN structure learning (see previous sub-section). We have used, $p_1$-model, initially proposed by Holland and Leinhardt \cite{p1model1981}, is a special class of exponential families of distributions, for this study that offer robust and flexible parametric models, which is used to evaluate the probability that a gene to be hub in the perturbation network inferred in previous sub-section. 

A STP BN model can be referred as a gene-gene relationship causal network with $g$ genes, which can be considered as a random variable $\textbf{U}$ over a space of $2^{g(g-1)}$ graphs. Let $\textbf{U=u}$ is the realization of $\textbf{U}$ and the binary outcome $u_{ij} = 1$ if $gene_i$ interacts with $gene_j$, or $u_{ij} = 0$ otherwise, then $\textbf{u}$ is a binary data matrix. Let $Pr(\textbf{u})$ be the probability function given by the following formula:

\begin{equation}
Pr(u)=Pr(\textbf{U}=\textbf{u})=\frac { 1 }{ \kappa \left( \boldsymbol{\theta}  \right)  } \exp\sum _{ p }^{  }{ { \boldsymbol{\theta}  }_{ p }{ z }_{ p }\left( \textbf{u} \right)  }
\end{equation}
where $z_p$(\textbf{u}) is the network statistic of type $p$, $\boldsymbol{\theta}_p$ is the parameter associated with $z_p$(\textbf{u}) and $\kappa(\boldsymbol{\theta})$ is the normalizing constant that ensures $Pr(\textbf{u})$ is a proper probability distribution (sums to 1 over all $\textbf{u}$ in $G$) \cite{Wasserman1996}. The parameter $\boldsymbol{\theta}$ is a vector of model parameters associated with network statistics and needs to be estimated. See \cite{$p_1$model1981} for further details. A major challenge in computing with the $p_1$-model is the calculation of $\kappa(\boldsymbol{\theta})$, for which the maximum likelihood estimation is intractable due to the large graph space cardinality. A technique called \textit{maximum pseudolikelihood estimation} \cite{Strauss1990} has been developed to address this issue, which employes MCMC sampling method such as Gibbs or Metropolis-Hastings sampling algorithms \cite{Snijders2002}   

In this study, we have considered the undirected version of $p_1$-model, which can be simplified by using only two Bernoulli variables $Y_{ij0}$ and $Y_{ij1}$ as follows:

\begin{displaymath}
   Y_{ijk} = \left\{
     \begin{array}{lr}
       1 & if \quad u_{ij} = k,\\
       0 & otherwise
     \end{array}
   \right.
\end{displaymath}
Now the $p_1$-model can be defined using the following two equations to predict the probability of an edge being present between $gene_i$ and $gene_j$:

\begin{equation}
\log\left\{ Pr\left( { Y }_{ ij1 }=1 \right)  \right\} ={ \lambda  }_{ ij }+\theta +{ \alpha  }_{ i }+\alpha _{ j }
\end{equation}

\begin{equation}
\log\left\{ Pr\left( { Y }_{ ij0 }=1 \right)  \right\} ={ \lambda  }_{ ij }
\end{equation}

for $i<j$, where $\theta$ is the global density parameter, and $\alpha$, which represents the \textit{sociality} of a gene to be connected in an undirected network. Note, $\lambda_{ij}$ is there to ensure $Pr\left( { Y }_{ ij0 }=1 \right) + Pr\left( { Y }_{ ij1 }=1 \right) = 1$.

\subsubsection*{\textbf{Bayesian model to infer perturbation drivers}}
Here, we have employed a fully Bayesian approach to infer the posteriori of parameters of the above $p_1$-model. For that purpose, MCMC sampling methods i.e. Gibbs sampling approach were adopted, which generate samples from the joint distribution of $P\left( {\cal M},{ \theta }|{\cal D} \right)$, where $\cal M$ is model under consideration, \textbf{$\theta$} is the set of model parameters to be inferred, and the \textbf{$\cal D$} is the data. Gibbs sampling iteratively samples on parameter at a time with full conditional distribution of the current and old values of other parameters. Our approach relies on a hierarchical Bayesian model where the model parameters depends on several hyper-parameters i.e., $\theta$ and $\alpha$, both assumed to be following a normal distribution, which in tern relies on 0 mean and standard deviation of $\sigma_\theta$ like the following:

\begin{equation}
\theta \sim {\cal N} \left( 0,{ { \sigma  }_{ \theta  } }^{ 2 } \right)
\end{equation}

\begin{equation}
\alpha \sim {\cal N} \left( 0,{ { \sigma  }_{ \alpha  } }^{ 2 } \right)
\end{equation}

Assuming $\tau={\sigma}^{-2}$, we can consider a gamma prior for both $\tau_\theta$ and $\tau_\alpha$ since, gamma is the conjugate prior of normal distribution:

\begin{equation}
{ \tau  }_{ \theta  }\sim Gamma\left( { a }_{ 0 },b_{ 0 } \right)
\end{equation}
\begin{equation}
{ \tau  }_{ \alpha  }\sim Gamma\left( { a }_{ 0 },b_{ 0 } \right)
\end{equation}
where we set $a_0$ = 0.001 and $b_0$ = 0.001 to make the prior for $\theta$ \textit{non-informative}, making its standard deviation wide enough to express large uncertainty \cite{Bulashevska2010}. To implement Gibbs sampling we have used JAGS (Just another Gibbs Sampler). We hypothesized that the perturbation driver genes would yield higher connectivity, i.e. reveal larger sociality score ($\alpha$ values) in a STP.






\section{Results}
\subsection*{\textbf{Differential Expression analysis and Functional enrichment test}}
\begin{figure}[hbt!]
  \centering
  \includegraphics[scale=0.85]{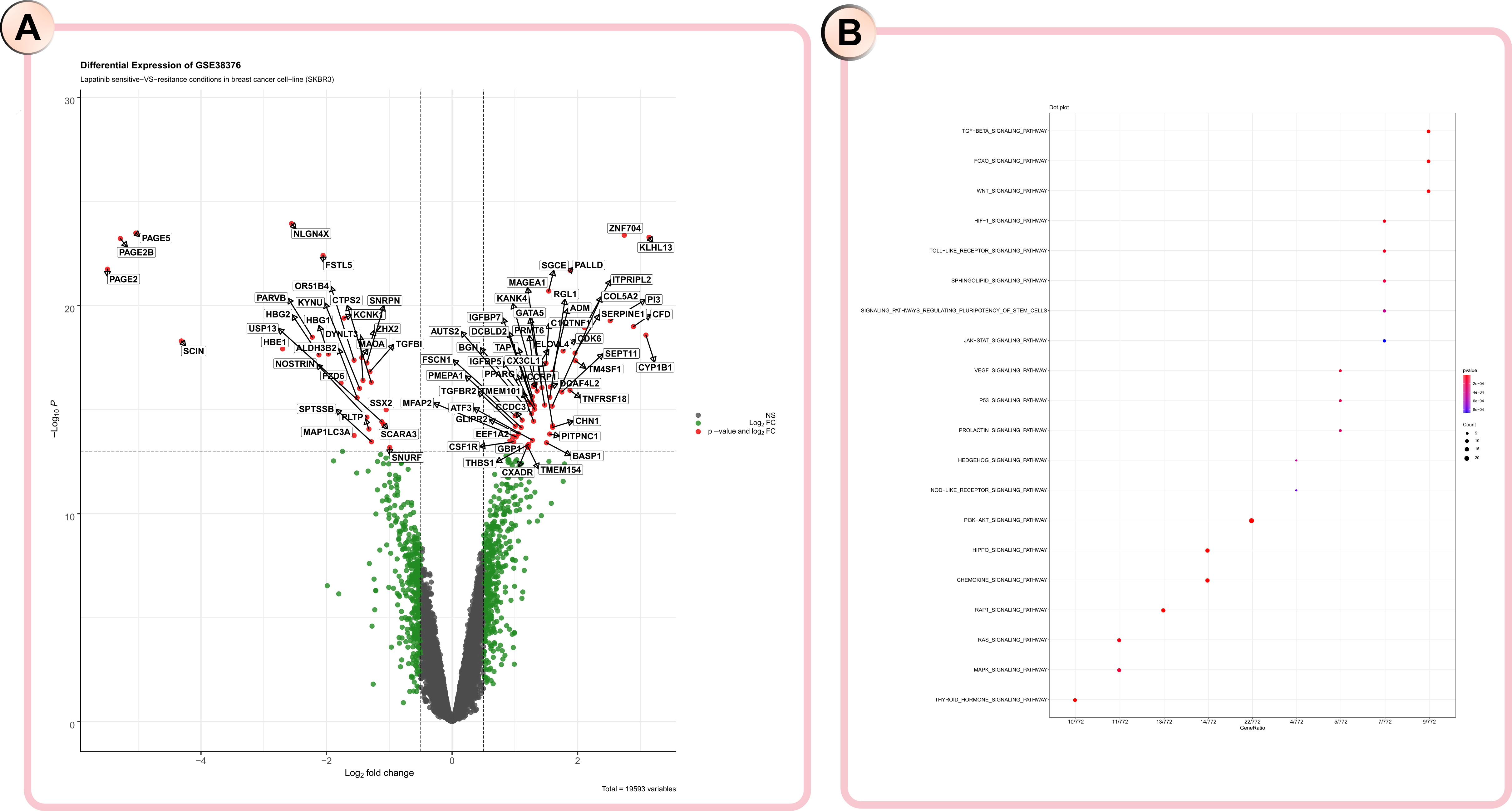}
  \caption{Differential expression (A) and function enrichment analyses (B) of resistant-vs-sensitive gene expression data in breast cancer cell-lines, SKBR3.}
  \label{fig_de_fn_enrich}
      \end{figure}

As a pilot experiment to observe the expression dynamics between case-vs-control studies, we conducted a differential expression analysis with lapatinib-sensitive and lapatinib-resistant gene expression from Breast cancer patients collected from Gene expression omnibus (accession number: GSE38376). Using limma R package with a threshold of 0.00001 for adjusted p-value (bonferroni corrected), we have found 512 differentially expressed genes (DEGs) [Additional File 1] as shown in Figure \ref{fig_de_fn_enrich}A. Next, to observed which signal transduction pathways (STPs) were enriched with the selected DEGs, we conducted a statistical over-representation analysis with a hyper-geometric test with \textit{phyper} function available in \textit{stat} R package. Using 45 KEGG signalling pathways, we have found 20 STPs that are significantly enrihced with the selected DEGs yeilding adjusted p-value $<$ 0.05, which are depicted in Figure \ref{fig_de_fn_enrich}B. As we observed the top significantly enriched STPs include PI3K-AKT signalling (adj. p-Value = $7.2\times10^{-9}$), Hippo signalling (adj. p-Value = $6.2\times10^{-8}$), Chemokine signalling (adj. p-Value = $1.02\times10^{-6}$), TGF-$\beta$ signaling (adj. p-Value = $3.21\times10^{-6}$), and Thyroid hormone signaling (adj. p-Value = $1.65\times10^{-5}$). As a case study of all our downstream experiments and analyses, i.e optimal STP structure learning and its perturbation driver identification, we have selected the TGF-$\beta$ signaling pathway (the number of nodes = 79) as it demonstrated significant enrichment with DEGs and reported to play significant role in breast cancer progression/resistance phenomenon via cross-talking with EGFR-mediated (lapatinib-targeted) signaling pathways \cite{Azad2015}.

\subsection*{\textbf{Novel MCMC methods for sampling Bayesian networks finds optimal STP as dys-regulated pathway in Lapatinib Resistance}}

\subsubsection*{\textbf{Data pre-processing}}
We hypothesized that the causal Bayesian network models may potentially reveal aberrant signalling activities given perturbation data from two-conditional dataset. At first, for a particular gene in a particular sample in the raw gene expression data for lapatinib-resistant cell-line (SKBR3) was standardized into z-score, based on the mean and standard deviation of expression of the all the samples in the corresponding gene in the lapatinib-sensitive conditions. This standardized score would measure how each resistant sample (case data) data is deviated from the control expression, which is in this case lapatinib-sensitive condition, thus revealing the perturbation measurements of resistant samples compared to the distribution of sensitive samples. Next, each expression values (i.e. z-score transformed) of lapatinib-resistant dataset were discretized in three levels based on thresholding ($>$ 1.5 or $<$ -1.5): \textit{over-expressed}, \textit{down-expressed}, or \textit{neutral} to model the data-driven Bayesian networks. 

\subsubsection*{\textbf{Optimal STP sampling for TGF-$\beta$ signalling using Neighbourhood Sampler, Hit-and-Run and Metropolis-Hastings algorithm}}
\begin{figure}[hbt!]
  \centering
  \includegraphics[scale=0.85]{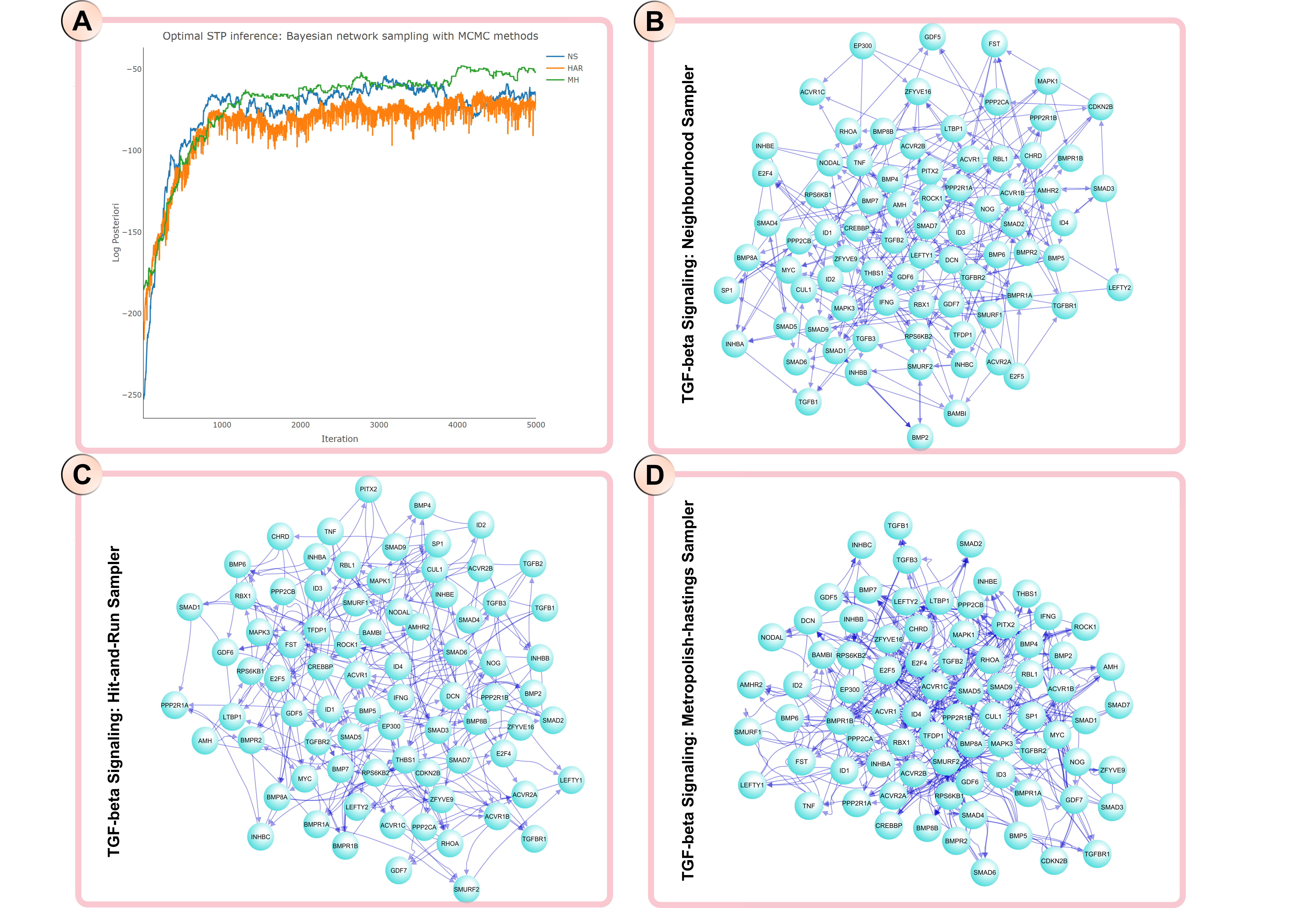}
  \caption{MCMC sampling and inferred perturbation BN model for TGF-$\beta$ signalling pathway using Neighbourhood sampler, Hit-and-Run sampler, and Metropolis-hasting sampler.}
  \label{fig_mcmc_bn}
      \end{figure}

In this study, we've considered a Bayesian network (BN) approach to model the signalling transduction pathways (STPs). With the standardized and discretized expression data for lapatinib-resistant cell-lines (SKBR3), we have conducted three MCMC sampling algorithms to search through a space of Bayesian network models of a STP of interest: i.e. TGF-$\beta$ signalling pathway. Those three MCMC sampling algorithms are: Neighbourhood sampler, Hit-and-Run sampler and Metropolis-hastings sampler. For each of these samplers, we chose the threshold of the number parents and the number of children is 4. Each sampling algorithm ran for 5000 sampling iteration and the burn-in iteration was considered as 3000, which indicates the latest 2000 iterations were used for actual sampling. Figure \ref{fig_mcmc_bn}A depicts the convergence of these three sampling algorithms, where Neighbourhood sampler (NS) starts to converges before Metropolis-Hastings (MH) and Hit-and-Run (HAR) samplers, but after the burn-in period (3000 iterations), all of them convergences were fairly observed. 

At each sampling iteration, each sampler picks a candidate BN models that achieves the best log-likelihood score among all others in their neighbourhood BN space. When finished, each sampler returns the BN that was sampled with highest probability of sampled (highest frequency out of the remaining 2000 sampling iterations) as the optimal STP model for the TGF-$\beta$ signalling pathway. Figure \ref{fig_mcmc_bn}B-D, depicts three TFG-$\beta$ BN models that were sampled from NS, HAR and MH samplers, respectively.


\subsection*{\textbf{Studying statistical properties of optimal STP reveals important genes that drive dys-regulation in lapatinib resistance}}

With the optimal STP model, we conducted experiment to observed the statistical aspect of the network formation with regards to their local structure formation. To that extent, we have used a fully Bayesian approach to infer the \textit{posteriori} of each node's \textit{sociality} information, which is analogous to it's importance as hub-gene in its interactome, and thus reveal its potentiality of being a gene that drive the perturbation in the signalling pathway. We have used the MCMC sampling approach again to infer the posteriori of network parameters of the $p_1$-model that represent the edge formation probability of an edge in the STP as a linear combination of several structure parameters, including $\alpha$, the sociality parameters of involving node-pairs. This MCMC sampling algorithm was conducted for each of the three inferred TGF-$\beta$ signalling pathways individually (see above) for 5000 iterations, 3000 of which were considered as the burn-in period. When completed, MCMC sampling with these three BN models of TGF-$\beta$ signalling pathway yielded 31, 33, and 22 social genes with of $\alpha$ values $>$ 0, respectively, that are listed along with their sensitive-vs-resistance log2 fold change in Table \ref{Tab1}. 

\begin{table}[htbp]
  \centering
  \caption{Perturbation drivers for three TGF-$\beta$ signalling pathway models identified by MCMC sampling}
    \begin{tabular}{lccccl} \toprule
    \multicolumn{1}{c}{\textbf{Gene}} & \textbf{From NS} & \textbf{From HAR} & \textbf{From MH} & \textbf{LogFC} & \textbf{Gene} \\
    \multicolumn{1}{c}{\textbf{Symbol}} & \textbf{sampling} & \textbf{sampling} & \textbf{sampling} &       & \textbf{Name} \\ \midrule
    NOG   & --    & 0.036803 & --    & 0.061517 & noggin \\
    THBS1 & 0.16256 & 0.152164 & --    & 1.221493 & thrombospondin 1 \\
    DCN   & --    & 0.028003 & --    & 0.10499 & decorin \\
    LEFTY1 & 0.166053 & --    & --    & 0.04226 & left-right determination factor 1 \\
    FST   & 0.054609 & 0.151098 & 0.081761 & 0.040524 & follistatin \\
    BMP2  & --    & 0.045136 & 0.082975 & -0.15355 & bone morphogenetic protein 2 \\
    BMP4  & 0.053143 & --    & 0.081372 & 0.04963 & bone morphogenetic protein 4 \\
    BMP5  & 0.057964 & --    & --    & -0.0029 & bone morphogenetic protein 5 \\
    BMP6  & 0.054911 & --    & --    & -0.01492 & bone morphogenetic protein 6 \\
    BMP7  & 0.051767 & 0.037539 & 0.177469 & -0.93311 & bone morphogenetic protein 7 \\
    BMP8B & --    & 0.0506 & --    & -0.0773 & bone morphogenetic protein 8b \\
    BMP8A & --    & 0.142381 & --    & 0.045929 & bone morphogenetic protein 8a \\
    GDF7  & 0.055338 & --    & --    & -0.18404 & growth differentiation factor 7 \\
    AMH   & 0.054537 & --    & 0.165438 & 0.095488 & anti-Mullerian hormone \\
    TGFB2 & 0.164959 & --    & --    & 0.706354 & transforming growth factor beta 2 \\
    INHBA & 0.054199 & 0.033856 & 0.077547 & -0.00524 & inhibin beta A subunit \\
    INHBB & --    & 0.030165 & --    & -0.38803 & inhibin beta B subunit \\
    NODAL & --    & 0.037555 & --    & -0.07754 & nodal growth differentiation factor \\
    BMPR2 & --    & 0.142948 & --    & 0.7048 & bone morphogenetic protein receptor type 2 \\
    TGFBR2 & --    & 0.160855 & --    & 1.005743 & transforming growth factor beta receptor 2 \\
    ACVR2B & 0.159379 & --    & 0.173425 & 0.133561 & activin A receptor type 2B \\
    BMPR1B & --    & 0.157393 & 0.082578 & 0.11865 & bone morphogenetic protein receptor type 1B \\
    ACVR1 & 0.16285 & 0.038792 & 0.173751 & 0.151194 & activin A receptor type 1 \\
    ACVR1B & 0.164295 & 0.035427 & --    & 0.086201 & activin A receptor type 1B \\
    BAMBI & --    & 0.037039 & 0.087164 & 0.350395 & BMP and activin membrane bound inhibitor \\
    SMAD1 & 0.050226 & --    & --    & 0.199748 & SMAD family member 1 \\
    SMAD5 & --    & 0.040064 & --    & -0.06311 & SMAD family member 5 \\
    SMAD9 & 0.057111 & --    & --    & 0.108667 & SMAD family member 9 \\
    SMAD2 & 0.161696 & --    & --    & 0.150392 & SMAD family member 2 \\
    SMAD3 & --    & 0.137773 & --    & 0.273339 & SMAD family member 3 \\
    SMAD4 & --    & --    & 0.173165 & 0.197391 & SMAD family member 4 \\
    SMAD6 & --    & 0.156279 & --    & -0.08563 & SMAD family member 6 \\
    SMAD7 & 0.055232 & --    & --    & 0.15036 & SMAD family member 7 \\
    SMURF1 & 0.056516 & 0.040997 & --    & 0.251961 & SMAD specific E3 ubiquitin protein ligase 1 \\
    ZFYVE9 & 0.053843 & 0.150341 & --    & 0.118893 & zinc finger FYVE-type containing 9 \\
    ZFYVE16 & 0.16138 & --    & 0.16913 & -0.05421 & zinc finger FYVE-type containing 16 \\
    ID1   & 0.166364 & --    & 0.175444 & 0.305308 & inhibitor of DNA binding 1, HLH protein \\
    ID3   & 0.167612 & 0.04118 & --    & -0.00622 & inhibitor of DNA binding 3, HLH protein \\
    ID4   & 0.04719 & --    & 0.173964 & 0.500908 & inhibitor of DNA binding 4, HLH protein \\
    RBL1  & --    & 0.043975 & --    & -0.25319 & RB transcriptional corepressor like 1 \\
    E2F5  & --    & 0.149304 & 0.172732 & 0.477958 & E2F transcription factor 5 \\
    TFD$p_1$ & --    & 0.15106 & --    & -0.17405 & transcription factor Dp-1 \\
    CREBBP & 0.055956 & 0.050031 & --    & 0.536597 & CREB binding protein \\
    EP300 & --    & 0.153432 & --    & 0.058004 & E1A binding protein p300 \\
    S$p_1$   & --    & 0.14603 & --    & -0.00254 & S$p_1$ transcription factor \\
    CDKN2B & 0.058374 & 0.149919 & --    & 1.095818 & cyclin dependent kinase inhibitor 2B \\
    PITX2 & --    & --    & 0.085195 & -0.60363 & paired like homeodomain 2 \\
    RBX1  & 0.16479 & --    & 0.083075 & -0.01859 & ring-box 1 \\
    CUL1  & --    & 0.149773 & --    & -0.01605 & cullin 1 \\
    MAPK1 & --    & --    & 0.084654 & 0.141796 & mitogen-activated protein kinase 1 \\
    MAPK3 & 0.165836 & 0.043168 & --    & 0.134163 & mitogen-activated protein kinase 3 \\
    IFNG  & 0.061818 & 0.145331 & 0.081417 & 0.085719 & interferon gamma \\
    RHOA  & --    & --    & 0.164908 & -0.26438 & ras homolog family member A \\
    PPP2CA & 0.059932 & --    & --    & 0.115046 & protein phosphatase 2 catalytic subunit alpha \\
    PPP2CB & --    & --    & 0.08517 & -0.38888 & protein phosphatase 2 catalytic subunit beta \\
    RPS6KB2 & 0.048165 & --    & --    & -0.18307 & ribosomal protein S6 kinase B2 \\
    ACVR1C & --    & --    & 0.075583 & 0.052312 & activin A receptor type 1C \\ \bottomrule
    \end{tabular}%
  \label{Tab1}%
\end{table}%


\subsection*{\textbf{Network analysis of aberrant STP}}
\begin{figure}[hbt!]
 \centering
 \includegraphics[scale=0.75]{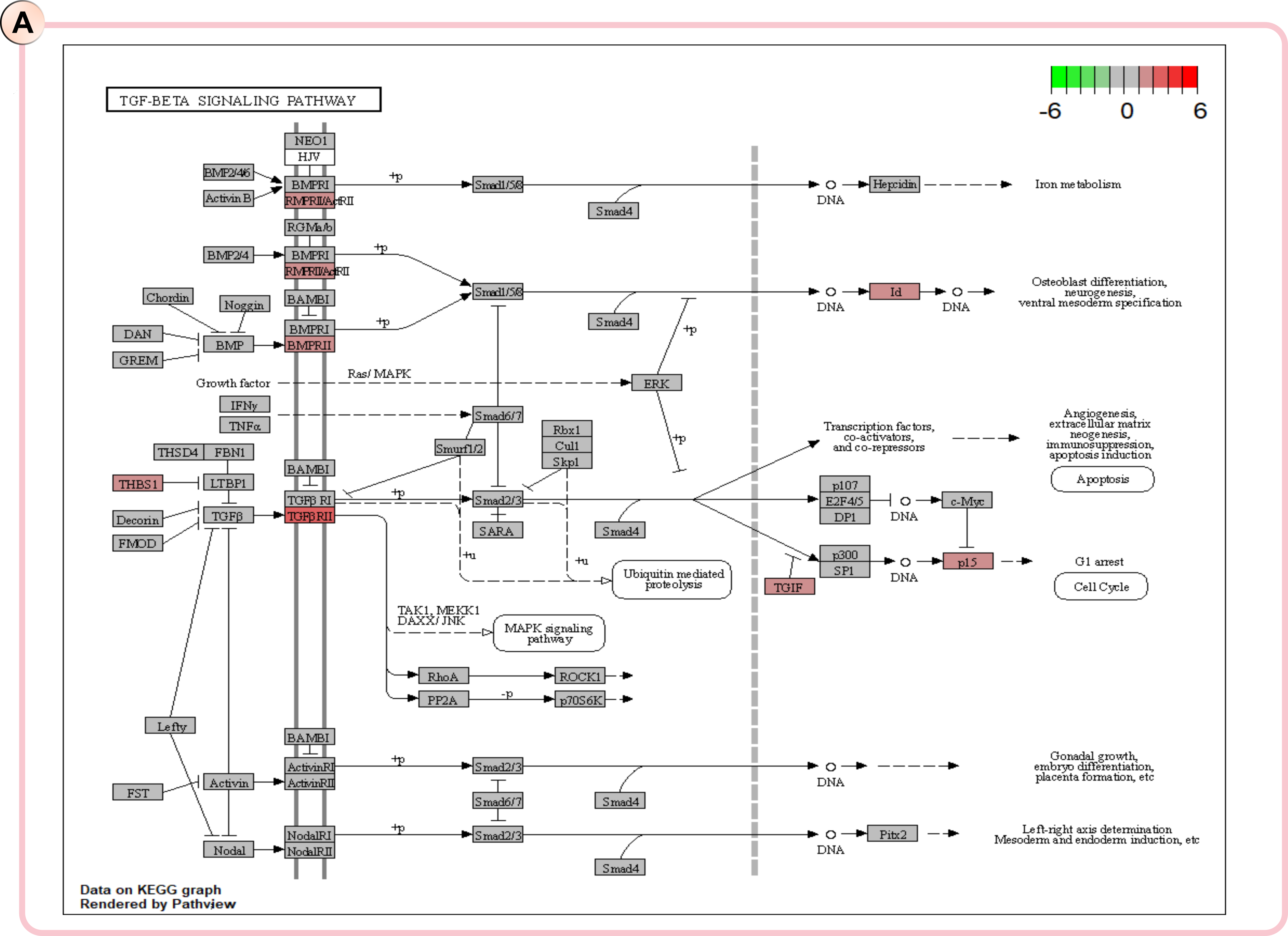}
  \caption{Perturbation detection of TGF-$\beta$ signalling pathway using PathView}
  \label{fig_tgf_pathview}
      \end{figure}

To observe the biological relevance of inferred driver genes in TGF-$\beta$ signalling pathways, we conducted GO term (Biological processes) enrichment using clueGO \cite{clueGO2019} plugin in Cytoscape. 
\begin{figure}[hbt!]
\centering
\includegraphics[scale=0.75]{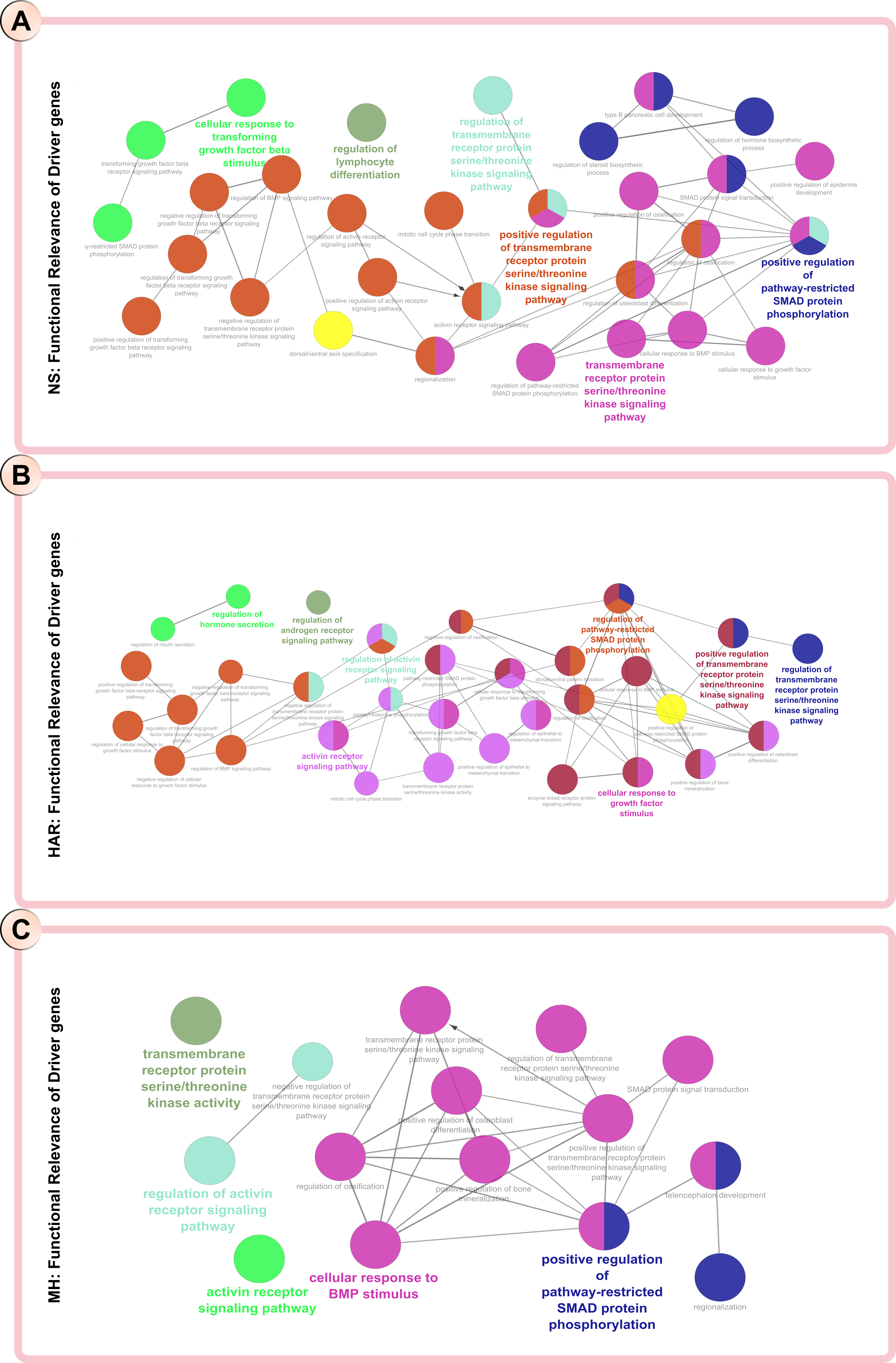}
  \caption{Functional enrichment of perturbation driver genes}
  \label{fig_fn_driver}
      \end{figure}
Figure \ref{fig_fn_driver}A-C depicts the significantly enriched GO terms, where positive regulation of pathway-restricted SMAD protein phosphorylation, activin receptor signalling, regulation of androgen receptor signalling, regulation of lymphocyte differentiation, and regulation of hormone secretion are observed. Using \textit{PathView} \cite{PathView2013} R package, we have also shown TFG-$\beta$ signalling pathway diagram overlayed with the constituent genes/proteins with their log2 fold change values in Figure \ref{fig_tgf_pathview}.

\subsection*{\textbf{Comparing with other methods in identifying aberrant STP}}

We have also experimented how our method performs in identifying pathway perturbation compared to other method. There are several approaches for finding perturbed pathways or functional terms. SPIA (Signaling pathway impact analysis) \cite{SPIA2009} is one of such approaches, which not only considers the functional enrichment tests but also the abnormal perturbation (based on pathway topology) of that pathway given a set of DE markers. SPIA have found that, both of those experiments are not necessarily dependent to each other,  i.e.  evidence from both of those tests signifies somewhat independent biological evidence.  Hence, SPIA conducts both (perturbation test + enrichment  tests)  and  statistically  combines  their  significance  and  provides  a combined significance level, namely \textit{PG}.  For mathematical details, please refer to the original article \cite{SPIA2009}. 
  \begin{figure}[hbt!]
  \centering
  \includegraphics[scale=0.75]{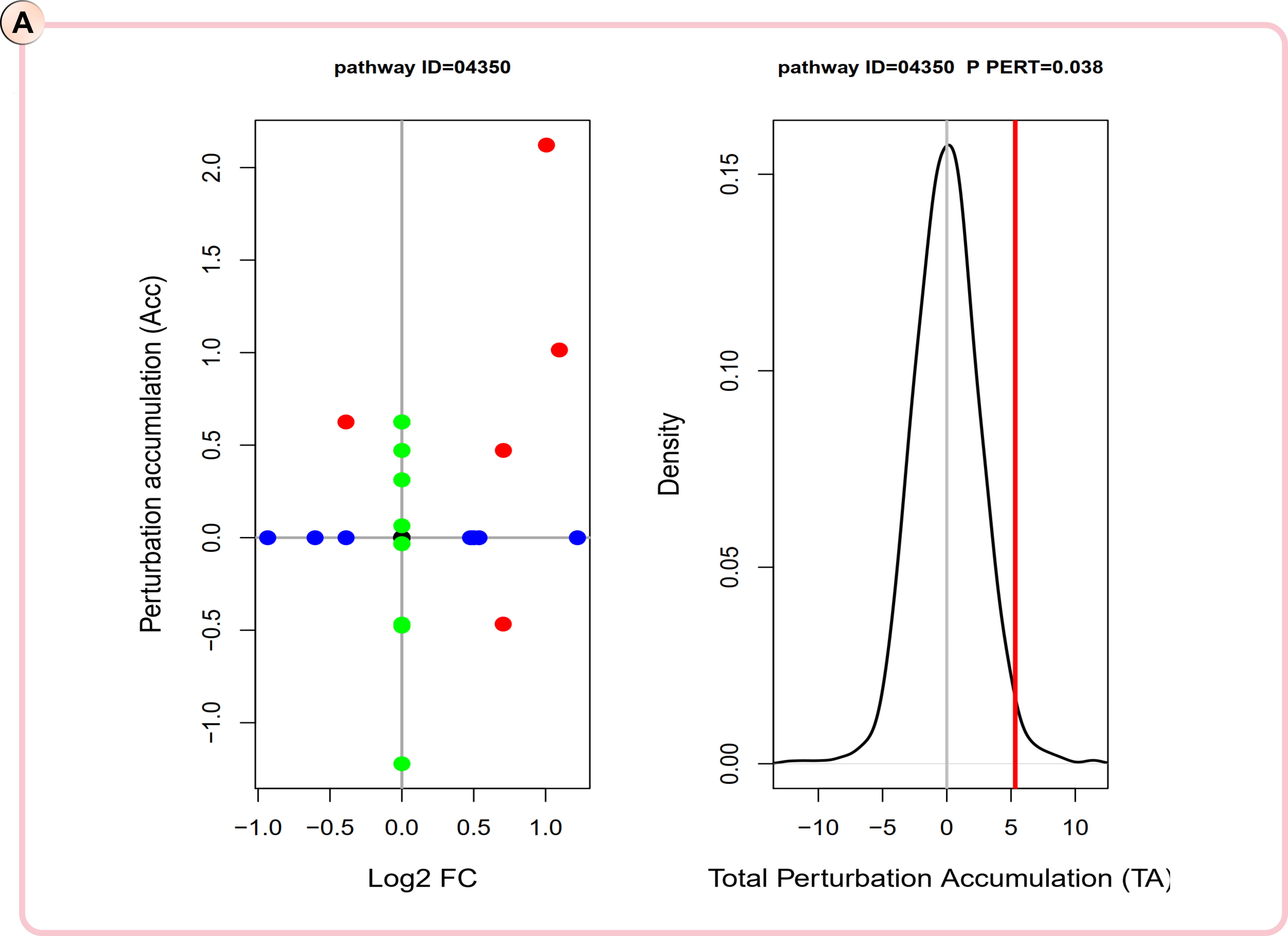}
  \caption{SPIA analyses to detect perturbation of TGF-$\beta$ signalling pathway}
  \label{fig_tgf_spia}
      \end{figure}
The left panel of Figure \ref{fig_tgf_spia} shows Log2 fold-change vs perturbation accumulation scores for TGF-$\beta$ signalling pathway (KEGG pathway ID = hsa04350) with their 9 enriched DEGs. Moreover the right panel shows the density plot of total perturbation accumulation scores and a probability of perturbation. As we can see from this plot that SPIA has detected TGF-$\beta$ signalling pathway as perturbed with the probability of perturbation, pPERT = 0.038 ($<$ 0.05), which is also coherent with our findings. 



\section*{Discussion}
In this study, namely `PathTurbEr', we provide a novel probabilistic approach of data-driven modeling of aberrant signal transduction pathways (STPs) via solving Bayesian Network structure learning problem, where each STP was modelled as a Bayesian Network. We adopted three MCMC sampling algorithm for this structure learning tasks, namely Neighbourhood sampler, Hit-and-Run sampler and Metropolis-Hastings sampler, where each of the sampling algorithm yielded an optimal STP BN model that best describes the aberration activities that are latent in two conditional studies, e.g. lapatinib resistant-vs-sensitive expression datasets. Note, this approach only models the aberrant activities in a signalling pathway of interest (e.g. TGF-$\beta$ signalling) by learning above two-conditional data, not just any data-driven structure of the signalling pathway. Next, we conducted MCMC sampling of the structural features of that those inferred STP to observe how the local feature can describe the overall structure formation probability by employing a fully Bayesian statistical modelling approach with $p_1$-model. This experiment yielded a set of important genes, namely social genes (analogous to hub genes in the interactome), in those aberrant BN models of STPs. Functional enrichment test and comparison with state-of-the-arts methods confirms the significant functional relevance of the inferred driver genes. This robustly summarizes the important contribution of our approach to this problem.

A set of bio-markers detected by differential expression analyses followed by the their functional pathway enrichment tests manifested several signalling pathways revealing perturbed expression in resistant-vs-sensitive conditions, including PI3K-AKT signalling, Hippo signalling, Chemokine signalling, TGF-$\beta$ signalling, and Thyroid hormone signalling. TGF-$\beta$ signalling has been reported to be contributing in resistance mechanism to targeted treatment in HER2-positive breast cancer \cite{Oliveras2012}. Komurov \textit{et al.} also reported TGF-$\beta$ signalling pathway as perturbed in resistant-vs-sensitive conditions in acquired lapatinib resistance in breast cancer cell-line, SKBR3 \cite{Komurov2012}. Tortora et al. previously reported that TGF-$\beta$ type I receptor activation is responsible for increased expression of HER ligands, and that mediates increased secretion of TGF-$\alpha$, amphiregulin, and heregulin, and that activates PI3K-AKT signalling pathway, which is also found to be perturbed in our analyses \cite{Tortora2011}. Moreover, our previous study of analysing drug-resistive cross-talks also revealed that TGF-$\beta$ signalling pathway cross-talks with EGFR/HER2 signalling pathway via activation of TFG-$\beta$ receptor, and thus provide a compensating route for resistant cells to survive despite continuous drug intake, i.e. lapatinib \cite{Azad2015}. 

Above evidence of TGF-$\beta$ signalling perturbation in resistant-vs-sensitive condition led us to investigate furthermore, which factors may drive this perturbation, i.e. if any bio-marker can be associated with this aberrant phenomenon. Therefore, we adopted a fully Bayesian statistical modelling approach in order to decipher important genes in a probabilistic manner. Unlike the frequentist/descriptive approach to identify hub genes as important genes in gene-gene relationship networks, probabilistic approaches adopt uncertainty in network formation, which is a intrinsic feature observed in biological networks, especially when derived from high-throughput experimental data \cite{Bulashevska2010}. Our hypothesis was that the statistically important genes in a given perturbation network structure in the context of connectivity, would indicate a key role underpinning the aberration. Results from this analysis revealed 31, 33, and 22 perturbation driver genes out of 79 constituent genes of three STP models of TGF-$\beta$ signalling yielded from Neighbourhood sampler, Hit-and-Run sampler, and Metropolis-Hastings sampler, respectively \ref{Tab1}. Out of these perturbation driver genes found by MCMC sampling, 9 genes, namely THBS1, TGFB2, TGFBR2, ID4, E3F5, CREBBP, CDKN2B, BMP7, and PITX2 were also identified as differentially expressed (first 7 genes were up-regulated whereas remaining were down-reguated) in resistant-vs-sensitive conditions in lapatinib resistance, in our independent differential expression analyses. 

Next, we aimed to observe the functional association of these putative genes driving the signalling perturbation by conducting an over-representation test with a set of GO terms (Biological process). We hypothesized that, the perturbation driver genes would be associated with many important biological functions triggered by TGF-$\beta$ receptor signalling that are relevant to breast cancer progression/resistance phenomenon. We have observed several biological functions including positive regulation of pathway-restricted SMAD protein phosphorylation, regulation of lymphocyte differentiation, positive regulation of transmembrane receptor protein serine/threonine kinase signaling, regulation of activin receptor signalling, cellular response to growth factor stimulus, regulation of tgf-beta receptor signalling, and cellular response to BMP stimulus. It has been reported that TGF-$\beta$-SMAD signalling plays significant role in EGFR/HER-positive breast cancer oncogenesis \cite{Sundqvist2020,Tarasewicz2012}. Chen \textit{et al.} recently reported that tumor infiltrating lymphocytes serves as critical pathological factor in predicting prognosis of breast cancer patients that are treated with anti-HER2 drugs, in our case which is lapatinib \cite{Chen2017}. The role of serine/threonine kinase signaling involving trans-membrane receptor proteins is well reported, i.e. it controls both cell proliferation and cell death in response to various stresses \cite{Manoharan2018}. Jeruss \textit{et al.} has also reported that regulation of both TGF-$\beta$ and activin signalling components are associated with advanced oncogenic progression in aggressive breast carcinoma \cite{Jeruss2003}. BMPs are found over-expressed in breast cancer patients, but results from Owens et al. suggested that the inhibition of BMP stimulated signalling may reduce breast cancer metastasis by targeting both the tumor and its surrounding micro-environments \cite{Owens2015}.

\section*{Conclusion}
Resistance to targeted therapies is a major obstacle in sustained and efficacious treatment plan for fighting against any cancer. Lapatinib, an EGFR/HER-2 dual inhibitor, although shown great initial promise in treating breast cancer patients, have ultimately been bypassed by the cancer-cells' alternate survival mechanism via compensatory pathways. Therefore, a root-cause analysis i.e. detection of bio-markers driving the pathway perturbation in resistant-vs-sensitive conditions with a robust framework is greatly needed to offer better therapeutic developments. Previous approaches mainly adopted descriptive methodologies with less emphasis on the stochastic phenomenon, which is very commonly exist in biological systems. Hence, `PathTurbEr' adopts statistical models (e.g. Bayesian networks) to detect aberrant signalling networks, followed by fully Bayesian statistical approach to identify the markers driving those perturbation. Our framework has been developed in a generalized way, therefore we hope that it can be applied to any case-vs-control expression studies, and analyse similar hypothesis for any given cancer datasets.

\section*{Competing interests}
  The authors declare that they have no competing interests.

\section*{Author's contributions}
    AKMA and SA conceived the idea. SA contributed to the MCMC sampling method design for Bayesian network inference from data. AKMA contributed the methodologies for statistical parameter sampling methodologies using MCMC methods. AKMA conducted the experiments, analysed the data and wrote the manuscript. SA contributed to the manuscript.

\section*{Additional Files}
\label{add_file_1}
  \subsection*{Additional file 1 --- 512 DEGs and their function enrichment results with 45 KEGG signalling pathways}

\bibliography{references}  

\begin{thebibliography}{10}

\bibitem{Sever2015}
R.~Sever and J.~S. Brugge.
\newblock {{S}ignal transduction in cancer}.
\newblock {\em Cold Spring Harb Perspect Med}, 5(4), Apr 2015.

\bibitem{Grumolato2017}
Luca Grumolato and Stuart~A. Aaronson.
\newblock {\em Aberrant Signaling Pathways in Cancer}, pages 1--8.
\newblock American Cancer Society, 2017.

\bibitem{Azad2017}
A.~K. Azad, A.~Lawen, and J.~M. Keith.
\newblock {{B}ayesian model of signal rewiring reveals mechanisms of gene
  dysregulation in acquired drug resistance in breast cancer}.
\newblock {\em PLoS One}, 12(3):e0173331, 2017.

\bibitem{Woolf2005}
P.~J. Woolf, W.~Prudhomme, L.~Daheron, G.~Q. Daley, and D.~A. Lauffenburger.
\newblock {{B}ayesian analysis of signaling networks governing embryonic stem
  cell fate decisions}.
\newblock {\em Bioinformatics}, 21(6):741--753, Mar 2005.

\bibitem{sachs2006}
Karen Sachs.
\newblock {\em Bayesian Network Models of Biological Signaling Pathways}.
\newblock PhD thesis, Massachusetts Institute of Technology, 2006.

\bibitem{Neapolitan2014}
R.~Neapolitan and X.~Jiang.
\newblock {{I}nferring {A}berrant {S}ignal {T}ransduction {P}athways in
  {O}varian {C}ancer from {T}{C}{G}{A} {D}ata}.
\newblock {\em Cancer Inform}, 13(Suppl 1):29--36, 2014.

\bibitem{NSpaper2016}
Salem~A. Alyami, A.~K.~M. Azad, and Jonathan~M. Keith.
\newblock {The neighborhood MCMC sampler for learning Bayesian networks}.
\newblock In Xudong Jiang, Guojian Chen, Genci Capi, and Chiharu Ishll,
  editors, {\em First International Workshop on Pattern Recognition}, volume
  10011, pages 326 -- 336. International Society for Optics and Photonics,
  SPIE, 2016.

\bibitem{barabasi204}
A.~L. Barab{\'a}si and Z.~N. Oltvai.
\newblock {{N}etwork biology: understanding the cell's functional
  organization}.
\newblock {\em Nat Rev Genet}, 5(2):101--113, Feb 2004.

\bibitem{Ernst2017}
M.~Ernst, Y.~Du, G.~Warsow, M.~Hamed, N.~Endlich, K.~Endlich, H.~Murua~Escobar,
  L.~M. Sklarz, S.~Sender, C.~Junghan?, S.~M?ller, G.~Fuellen, and
  S.~Struckmann.
\newblock {{F}ocus{H}euristics - expression-data-driven network optimization
  and disease gene prediction}.
\newblock {\em Sci Rep}, 7:42638, 02 2017.

\bibitem{wasserman_robins_2005}
Stanley Wasserman and Garry Robins.
\newblock {\em An Introduction to Random Graphs, Dependence Graphs, and p*},
  page 148–161.
\newblock Structural Analysis in the Social Sciences. Cambridge University
  Press, 2005.

\bibitem{Bulashevska2010}
S.~Bulashevska, A.~Bulashevska, and R.~Eils.
\newblock {{B}ayesian statistical modelling of human protein interaction
  network incorporating protein disorder information}.
\newblock {\em BMC Bioinformatics}, 11:46, 2010.

\bibitem{Komurov2012}
K.~Komurov, J.~T. Tseng, M.~Muller, E.~G. Seviour, T.~J. Moss, L.~Yang,
  D.~Nagrath, and P.~T. Ram.
\newblock {{T}he glucose-deprivation network counteracts lapatinib-induced
  toxicity in resistant {E}rb{B}2-positive breast cancer cells}.
\newblock {\em Mol Syst Biol}, 8:596, 2012.

\bibitem{KEGG2002}
M.~Kanehisa.
\newblock {{T}he {K}{E}{G}{G} database}.
\newblock {\em Novartis Found Symp}, 247:91--101, 2002.

\bibitem{limma2015}
M.~E. Ritchie, B.~Phipson, D.~Wu, Y.~Hu, C.~W. Law, W.~Shi, and G.~K. Smyth.
\newblock {limma powers differential expression analyses for
  {R}{N}{A}-sequencing and microarray studies}.
\newblock {\em Nucleic Acids Res}, 43(7):e47, Apr 2015.

\bibitem{Salemthesis2017}
{SALEM ALI S ALYAMI}.
\newblock Markov chain monte carlo methods for bayesian network inference, with
  applications in systems biology.
\newblock 2017.

\bibitem{p1model1981}
Paul~W. Holland and Samuel Leinhardt.
\newblock An exponential family of probability distributions for directed
  graphs.
\newblock {\em Journal of the American Statistical Association},
  76(373):33--50, 1981.

\bibitem{Wasserman1996}
Stanley Wasserman and Philippa Pattison.
\newblock Logit models and logistic regressions for social networks: I. an
  introduction to markov graphs andp.
\newblock {\em Psychometrika}, 61(3):401--425, 1996.

\bibitem{Strauss1990}
D~Strauss and M~Ikeda.
\newblock Pseudolikelihood estimation for social networks.
\newblock {\em Journal of the American Statistical Association},
  85(409):204--212, March 1990.

\bibitem{Snijders2002}
Tom A.~B. Snijders.
\newblock Markov chain monte carlo estimation of exponential random graph
  models.
\newblock {\em Journal of Social Structure}, 3, 2002.

\bibitem{Azad2015}
A.~K. Azad, A.~Lawen, and J.~M. Keith.
\newblock {{P}rediction of signaling cross-talks contributing to acquired drug
  resistance in breast cancer cells by {B}ayesian statistical modeling}.
\newblock {\em BMC Syst Biol}, 9:2, Jan 2015.

\bibitem{clueGO2019}
B.~Mlecnik, J.~Galon, and G.~Bindea.
\newblock {{A}utomated exploration of gene ontology term and pathway networks
  with {C}lue{G}{O}-{R}{E}{S}{T}}.
\newblock {\em Bioinformatics}, 35(19):3864--3866, 10 2019.

\bibitem{PathView2013}
W.~Luo and C.~Brouwer.
\newblock {{P}athview: an {R}/{B}ioconductor package for pathway-based data
  integration and visualization}.
\newblock {\em Bioinformatics}, 29(14):1830--1831, Jul 2013.

\bibitem{SPIA2009}
A.~L. Tarca, S.~Draghici, P.~Khatri, S.~S. Hassan, P.~Mittal, J.~S. Kim, C.~J.
  Kim, J.~P. Kusanovic, and R.~Romero.
\newblock {{A} novel signaling pathway impact analysis}.
\newblock {\em Bioinformatics}, 25(1):75--82, Jan 2009.

\bibitem{Oliveras2012}
C.~Oliveras-Ferraros, B.~Corominas-Faja, S.~Cuf?, A.~Vazquez-Martin,
  B.~Martin-Castillo, J.~M. Iglesias, E.~L?pez-Bonet, ?.~G. Martin, and J.~A.
  Menendez.
\newblock {{E}pithelial-to-mesenchymal transition ({E}{M}{T}) confers primary
  resistance to trastuzumab ({H}erceptin)}.
\newblock {\em Cell Cycle}, 11(21):4020--4032, Nov 2012.

\bibitem{Tortora2011}
Giampaolo Tortora.
\newblock {Mechanisms of Resistance to HER2 Target Therapy}.
\newblock {\em JNCI Monographs}, 2011(43):95--98, 10 2011.

\bibitem{Sundqvist2020}
A.~Sundqvist, E.~Vasilaki, O.~Voytyuk, Y.~Bai, M.~Morikawa, A.~Moustakas,
  K.~Miyazono, C.~H. Heldin, P.~Ten~Dijke, and H.~van Dam.
\newblock {{T}{G}{F}-beta and {E}{G}{F} signaling orchestrates the {A}{P}-1-
  and p63 transcriptional regulation of breast cancer invasiveness}.
\newblock {\em Oncogene}, 39(22):4436--4449, 05 2020.

\bibitem{Tarasewicz2012}
E.~Tarasewicz and J.~S. Jeruss.
\newblock {{P}hospho-specific {S}mad3 signaling: impact on breast oncogenesis}.
\newblock {\em Cell Cycle}, 11(13):2443--2451, Jul 2012.

\bibitem{Chen2017}
T.~H. Chen, Y.~C. Zhang, Y.~T. Tan, X.~An, C.~Xue, Y.~F. Deng, W.~Yang,
  X.~Yuan, and Y.~X. Shi.
\newblock {{T}umor-infiltrating lymphocytes predict prognosis of breast cancer
  patients treated with anti-{H}er-2 therapy}.
\newblock {\em Oncotarget}, 8(3):5219--5232, Jan 2017.

\bibitem{Manoharan2018}
R.~Manoharan, H.~A. Seong, and H.~Ha.
\newblock {{D}ual {R}oles of {S}erine-{T}hreonine {K}inase
  {R}eceptor-{A}ssociated {P}rotein ({S}{T}{R}{A}{P}) in {R}edox-{S}ensitive
  {S}ignaling {P}athways {R}elated to {C}ancer {D}evelopment}.
\newblock {\em Oxid Med Cell Longev}, 2018:5241524, 2018.

\bibitem{Jeruss2003}
Jacqueline~S. Jeruss, Charles~D. Sturgis, Alfred~W. Rademaker, and Teresa~K.
  Woodruff.
\newblock Down-regulation of activin, activin receptors, and smads in
  high-grade breast cancer.
\newblock {\em Cancer Research}, 63(13):3783--3790, 2003.

\bibitem{Owens2015}
P.~Owens, M.~W. Pickup, S.~V. Novitskiy, J.~M. Giltnane, A.~E. Gorska, C.~R.
  Hopkins, C.~C. Hong, and H.~L. Moses.
\newblock {{I}nhibition of {B}{M}{P} signaling suppresses metastasis in mammary
  cancer}.
\newblock {\em Oncogene}, 34(19):2437--2449, May 2015.

\end{thebibliography}

\bibliographystyle{unsrt}  

\end{document}